\documentclass[12pt]{iopart}

\usepackage{iopams}
\usepackage{graphicx}
\usepackage[percent]{overpic}
\usepackage{xcolor}
\usepackage{hyperref}
\usepackage{verbatim}
\usepackage{cite}

\begin{document}
\let \vec \mathbf

\renewcommand{\[}{\begin{equation}}
\renewcommand{\]}{\end{equation}}

\title[Hydrodynamic PFC approach to interfaces, dislocations, and multi-grain networks]{Hydrodynamic phase field crystal approach to interfaces, dislocations, and multi-grain networks}

\author{Vidar Skogvoll$^{1}$, Marco Salvalaglio$^{2,3}$, Luiza Angheluta$^{1}$}
\address{$^1$PoreLab, Njord Centre, Department of Physics, University of Oslo, P. O. Box 1048, 0316 Oslo, Norway.}
\address{$^2$Institute of Scientific Computing, TU Dresden, 01062 Dresden, Germany.}
\address{$^3$Dresden Center for Computational Materials Science, TU Dresden, 01062 Dresden, Germany.}


\begin{abstract}
We derive a phase field crystal model that couples the diffusive evolution of a microscopic structure with the fast dynamics of a macroscopic velocity field, explicitly accounting for the relaxation of elastic excitations.
This model captures better than previous formulations the dynamics of complex interfaces and dislocations in single crystals as well as grain boundary migration in poly-crystals where the long-range elastic field is properly relaxed.
The proposed model features a diffusivity that depends non-linearly on the local phase.
It induces more localized interfaces between a disordered phase (liquid-like) and an ordered phase, e.g., stripes or crystal lattices. For stripes, the interface dynamics are shown to be strongly anisotropic.
We also show that the model is able to evolve the classical PFC at mechanical equilibrium. However, in contrast to previous approaches, it is not restricted to a single-crystal configuration or small distortions from a fixed reference lattice.
To showcase the capabilities of this approach, we consider a few examples, from the annihilation of dislocation loops in a single crystal at mechanical equilibrium to the relaxation of a microstructure including crystalline domains with different orientations and grain boundaries.
During the self-annihilation of a mixed type dislocation loop (i.e., not shear or prismatic), long-range elastic effects cause the loop to move out of plane before the annihilation event.
\end{abstract}

\section{Introduction}
The phase field crystal model (PFC) is a continuum field theory describing crystalline systems at diffusive time scales while naturally incorporating elastic and plastic deformation \cite{Elder2002,Elder2004,Emmerich2012}.
It encodes key features for multiscale investigations of mechanical properties in crystals, such as the role played by dislocation dynamics and grain boundary migration in the dissipation of the elastic energy and control of large-scale deformations.
This model is based on the Swift-Hohenberg free-energy functional, which is minimized by a periodic order parameter.
Typically, a conservative, over-damped dynamic describes the evolution of out-of-equilibrium configurations for this order parameter which then represents a microscopic density field averaged over vibrational time scales.
Importantly, this field describes defects of the periodic lattice, meaning that the density field and its dynamic can describe slip planes, nucleation, dislocation interaction and motion, as well as the patterns formed by their collective behavior.
The PFC model contains a limited number of tunable parameters in its original formulation, usually allowing for a qualitative description of the phenomena mentioned above.
However, formulations featuring extended sets of parameters and coupling to additional physical contributions have been proposed, enabling quantitative descriptions \cite{Greenwood2010,Seymour2015,Hirvonen2016}.
Elasticity within the PFC model arises naturally when considering small deformations of the density field. Also, the characterization of elastic fields at continuum length scales through coarse-graining procedures has been widely discussed in the literature \cite{Spatschek2010,heinonenPhasefieldcrystalModelsMechanical2014,skaugenDislocationDynamicsCrystal2018,skaugenSeparationElasticPlastic2018,salvalaglioClosingTheGapNPJ2019,skogvollStressOrderedSystems2021}. However, the consistent coupling of the diffusive dynamics with the additional time scale associated with elastic relaxation within the PFC model is still an open problem.
Indeed, it has been addressed to some extent by different formulations focusing either on some specific limits or on derived methods, which can be all traced back to two general approaches.

The first approach eliminates the long-wavelength elastic waves through an additional mechanical equilibrium constraint imposed on the diffusive PFC dynamics. This idea has been first developed by Heinonen et al.~\cite{heinonenPhasefieldcrystalModelsMechanical2014} in the amplitude expansion formulation of the PFC model (APFC) \cite{Goldenfeld2005,Athreya2006,salvalaglio2022coarse}, describing the crystal structure through slowly-varying (complex) amplitudes of a small set of Fourier modes approximating the periodic density field.
It consists of an operator-splitting method that evolves the elastic excitations independently between each diffusive time step.
A limitation of this formulation is that the APFC model cannot handle large lattice rotations and thus cannot be applied to study poly-crystalline systems or highly deformed crystals.
This can be resolved by imposing the instantaneous mechanical equilibrium condition directly onto the diffusive evolution of the phase field crystal density~\cite{skaugenSeparationElasticPlastic2018,salvalaglioCoarsegrainedPhasefieldCrystal2020,skogvollPhaseFieldCrystal2022}.
The resulting model allows addressing problems where the time scale of elastic interactions is assumed to be much faster than the dissipative evolution of, e.g., grain boundaries or dislocation motion.
However, neglecting inertial elastic effects is a singular limit analogous to the Stokes or viscous-dominated flows in fluid hydrodynamics.
This limit may fail to quantitatively describe certain fast events such as defect annihilation or nucleation, where the interaction between elastic waves and dislocation motion is expected to become relevant.

The other approach considers these inertial elastic effects and is based on coupling the diffusive PFC dynamics with a momentum balance law corresponding to an appropriate velocity field.
A first attempt at including elastic wave effects led to an additional second-order temporal derivative in the PFC dynamics \cite{stefanovicPhasefieldCrystalsElastic2006}.
This model features an additional time scale and qualitatively captures some short-range elastic behavior.
However, it has been shown that it fails to reproduce the correct phonon dispersion and describe large-scale elastic perturbations ~\cite{majaniemiDissipativePhenomenaAcoustic2007}.
In later attempts, a velocity field was introduced, and its flow equation was coupled directly with the overdamped dynamic of the original PFC model.
An initial problem with this approach is that the microscopic nature of the PFC density can lead to large and unphysical flows.
Using the already coarse-grained APFC formulation, a consistent treatment has been proposed in Ref. \cite{heinonenConsistentHydrodynamicsPhase2016}.
This approach gives the correct dispersion relation for the long-wavelength elastic waves in the density field, but it is valid for small deformation and rotations only.
Employing the full atomic resolution, but introducing the notion of coarse-graining through the use of Fourier filters, Ref. \cite{tothNonlinearHydrodynamicTheory2013} proposed a coarse-graining operation of the density and momentum, so that dissipation of the velocity field was handled in the long-range limit.
However, in its original form, the model lacked the classical first-order dissipative term used in phase field modeling.
This term was reintroduced into the model for computational efficiency in Ref. \cite{podmaniczkyNucleationPostNucleationGrowth2021} as a phenomenological damping coefficient.
Damping coefficients, however, typically imply the interaction with a colloid or surrounding media and are hard to justify when considering a single component system.

In this paper, we show a consistent derivation for the coupling of the PFC dynamics with a macroscopic velocity field using linear response theory of relaxation to equilibrium.
The derived equations of motion are similar to the model proposed in Ref.~\cite{podmaniczkyNucleationPostNucleationGrowth2021}. However, we provide a different physical interpretation of the source of free energy dissipation, which is related to the small-scale variations (microscopic structures) of the density field.
Postulating that mass diffusion happens due to the emergence of such microscopic structures, we show that the structure tensor appears as a natural choice of spatially dependent mobility in the model, which is identically zero in the liquid phase so that the model reduces to classical hydrodynamics in this limit.
Like in Ref. \cite{tothNonlinearHydrodynamicTheory2013}, the separation of length scales is done by introducing a spatial coarse-graining and showing that the momentum density couples only to long-wavelength variations in the microscopic density field.
A microscopic crystalline structure sustains shear deformation, and the corresponding stress fields derived from this model are in excellent agreement with the ones obtained in the instantaneous mechanical equilibrium approximation from Ref.~\cite{skogvollPhaseFieldCrystal2022}.
We also show that our model reduces in some limits to the consistent (APFC) hydrodynamic model from Ref.~\cite{heinonenConsistentHydrodynamicsPhase2016}.
However, it is more versatile because it has no constraints on the lattice orientation  and, therefore, is more suited to study problems involving grain boundaries, lattice rotations, and large deformations.
We compare the propagation of a solidification front of our model to that predicted by the classical PFC dynamics, which shows that the solid-liquid interface is sharper with the spatially dependent mobility tensor than without, and predicts anisotropic growth rates in some cases.
We show that for suitable model parameters, our hydrodynamic model reproduces very well the same features observed using instead the instantaneous mechanical equilibrium constraint introduced in Ref. \cite{skogvollPhaseFieldCrystal2022}.
Moreover, as proof of concept, we highlight with a few examples that our model can capture very well dislocation dynamics in different slip planes, as well as their interactions with elastic waves and their pile-ups into grain boundary networks.

\section{Hydrodynamic model with microscopic diffusion}

\subsection{Classical hydrodynamics}
\label{sec:classical_hydrodynamics}
Classical dissipative hydrodynamics can be derived using linear response theory for the relaxation to equilibrium starting from a free energy given by
\[
F[\rho] = \int d\vec r \left (f(\rho) + \frac{1}{2} \rho  |\vec v|^2\right),
\label{eq:free_energy_coarse_grained}
\]
where $f(\rho)$ is some local free energy density determined by a density field $\rho \equiv \rho(\vec r, t)$ and the last term is the kinetic energy density corresponding to a velocity field $\vec v \equiv \vec v(\vec r,t)$.
We examine its time derivative, which depends on both the evolution of the density and the velocity as
\[
\partial_t F =  \int d\vec r  \left ((\mu_{\rm c}  +\frac{1}{2}  |\vec v|^2) \partial_t \rho + \rho \vec v \partial_t \vec v \right ),
\]
where $\mu_{\rm c} = \partial f/\partial \rho$ is the chemical potential.
The evolutions of mass and momentum follow the generic conservation laws in the form
\[
\partial_t \rho + \nabla \cdot (\rho \vec v)= 0,
\label{eq:liquid_conservation_of_mass}
\]
and
\begin{equation}\label{eq:momentum}
    \partial_t (\rho\vec v) + \nabla \cdot (\rho \vec v\vec v)= \nabla \cdot \sigma ,
\end{equation}
where $\sigma$ is the stress tensor.
By inserting the continuity equation into the momentum equation, we can write the latter equivalently as
\[
 \rho D_t  \vec v = \nabla \cdot  \sigma ,
\]
where $D_t$ is the material derivative $D_t X = \partial_t X + (\vec v \cdot \nabla) X$.
Combining these dynamical equations with the time evolution of $F$, we find that
\[
\begin{array}{ll}
\partial_t F
 &= \int d\vec r \left [
-\mu_{\rm c} \nabla \cdot (\rho \vec v) + \vec v \cdot (\nabla \cdot \sigma)
\right ].
\end{array}
\]
By repeated integration by parts and ignoring boundary terms, we get
\[
\partial_t F = \int d\vec r \left (
-( \sigma  - \sigma^{(\rho)}):\nabla   \vec v
\right )
,
\]
where we have used $\nabla \cdot \sigma^{(\rho)} = -\rho \nabla \mu_{\rm c}$ for $\sigma^{(\rho)}_{ij} = -P\delta_{ij}$ where $P = -(f-\mu_{\rm c}\rho)$ is the pressure of an isotropic fluid \cite{chaikinPrinciplesCondensedMatter1995,skogvollStressOrderedSystems2021}.
From the second law of thermodynamics, we require that $\partial_t F<\dot W$, where $\dot W$ is the power or the rate of mechanical work done on the system by external forces, given by
\[
\dot W = \int  d\vec r [\vec f^{\rm (ext)} \cdot \vec v] =  -\int d\vec r [\sigma^{\rm (ext)} : \nabla   \vec v]
\]
where $\vec f^{(\rm{ext})} \equiv \nabla \cdot \sigma^{(\rm{ext})}$ is the external body force density and $\sigma^{(\rm{ext})}$ the external stress tensor.
This is achieved by postulating a linear response $\sigma - \sigma^{(\rm{ext})} - \sigma^{(\rho)} = \mathcal D : \nabla   \vec v$, where the coefficient $\mathcal D$ is a positive definite rank-four tensor and $ $ represents the tensor product.
We are using tensor notation throughout, where the dot product ($\cdot$) is a contraction over the last index, e.g., $(\nabla  \cdot \sigma)_i = \nabla_j \sigma_{ij}$, and the double dot product ($:$) is a contraction over the last two indices, e.g., $(\mathcal D : \nabla   \vec v)_{ij} = \mathcal D_{ijkl} \nabla_k v_l$.
The stress tensor $\sigma$ thus has three contributions: a reversible part $\sigma^{(\rho)}$ dependent on the state of the system, a dissipative part given by an irreversible stress $\sigma^{(\rm{irr})} = \mathcal D : \nabla \vec v$, and a contribution from the external stress $\sigma^{\rm (ext)}$.
The momentum balance law reduces to
\[
\rho D_t \vec v
=-\nabla P + \vec f^{(\rm{ext})} + \nabla \cdot ( \mathcal D : \nabla  \vec v).
\]
A typical choice of the dissipation coefficient tensor is the isotropic form determined by the kinematic viscosity $\nu$ and this leads to the classical Navier-Stokes equation
\[
\rho D_t \vec v = -\nabla P + \vec f^{(\rm{ext})} + \nu \nabla^2\vec v,
\]
which, together with the density conservation law, describes the evolution of compressible, homogeneous, and isotropic fluids.

\subsection{Hydrodynamic model with microscopic structures}

\begin{figure}[htp]
    \centering
    \includegraphics[]{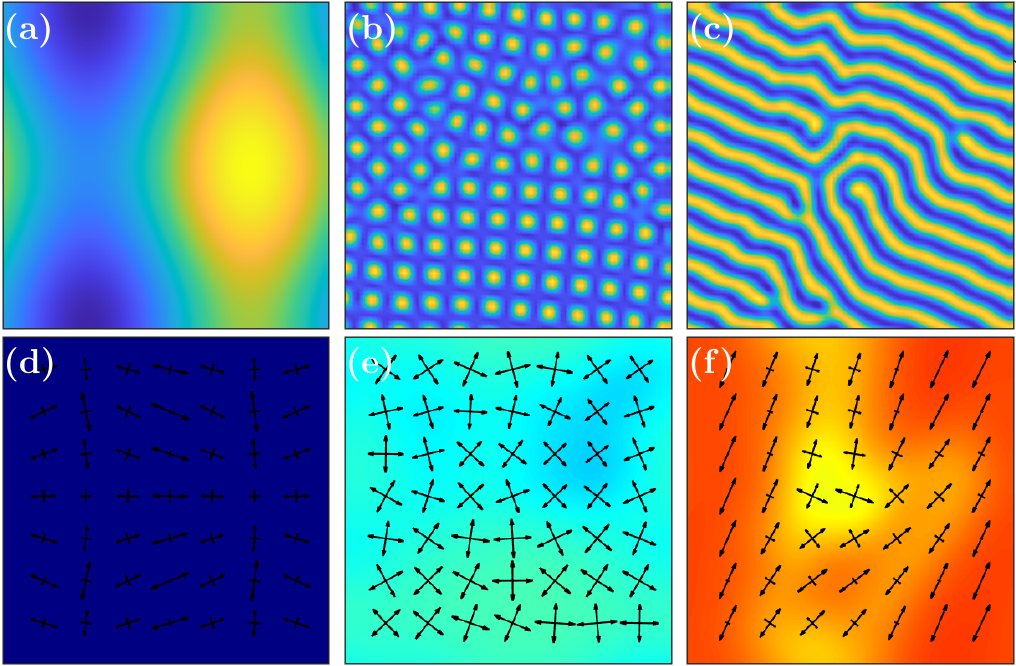}
    \caption{Examples of microscopic structures (of length scale $w$) in a density field that can form with a non-local free energy density.
    These plots were produced by minimizing the PFC free energy, Eq.~(\ref{eq:F_PFC}) with constant average density $\psi_0$ from a random initial state (quenching) using $\mathcal L = \mathcal L_{2M}$ and parameters $(q_0,\texttt B_0,\texttt t,\texttt v)=(1,1,0,1)$ to favor the emergence of (a) a slowly varying liquid phase ($(\psi_0,\Delta \texttt B_0^x) = (0.2,0.2)$), (b) a crystal phase with square symmetry ($(\psi_0,\Delta \texttt B_0^x) = (-0.3,-0.3)$), and (c) a striped phase  ($(\psi_0,\Delta \texttt B_0^x) = (0.0,-0.3)$).
    (d-f) The structure tensors $\mathcal S^{(\tilde \rho)}$ for the density profiles in (a-c), respectively.
    The tensors are represented by the directions of their eigenvectors $(\vec e_1,\vec e_2)$ and scaled according to the eigenvalues $(\lambda_1,\lambda_2)$.
    The color map in (e-f) shows $\sqrt{\lambda_1^2 + \lambda_2^2}$ with the same colorbar ranging from $0$ (dark blue) to $0.24$ (red) in dimensionless units.
    Diffusion in Eq.~\ref{eq:diffusion_along_eigenvectors} will occur along the directions given in (d-f).}
    \label{fig:MicroStructures}
\end{figure}

In this section, we show how to couple in a self-consistent way the diffusive transport of a microscopic density field with a macroscopic velocity field. Whereas the free energy in Eq.~(\ref{eq:free_energy_coarse_grained}) describes a homogeneous and isotropic material, particular microscopic structures can be incorporated by including additional non-local terms in the free energy density.
In the simplest formulation and the spirit of Landau's theory, the non-local terms can be expanded in terms of gradients of the density field. This gradient expansion introduces a microscopic length $w$ representative of the small-scale variations.
Figure \ref{fig:MicroStructures} (a-c) shows a few examples of microscopic structures that can be represented through a gradient-based free energy density.
Henceforth, quantities that also vary on this microscopic scale will be denoted by a $\sim$, e.g.,  $\tilde \rho$ and $\tilde f$ refer to the rapidly-varying mass density and the free energy density, respectively.
To generalize Eq.~(\ref{eq:free_energy_coarse_grained}), we assume that only the long-wavelength velocity $\vec v$ contributes to large-scale transport, while the small-scale flows typically results in dissipative small-scale transport.
Thus,  Eq.~(\ref{eq:free_energy_coarse_grained}) reads as
\[
F[\tilde \rho] = \int d\vec r \left (\tilde f(\tilde \rho,\nabla \tilde \rho, ...) + \frac{1}{2}\tilde \rho  |\vec v|^2\right) \equiv F_{\tilde \rho}[\tilde \rho] + \int d\vec r \frac{1}{2}\tilde \rho  |\vec v|^2,
\]
and the time derivative is then
\[
\partial_t F = \int d\vec r  \left ((\tilde \mu_{\rm c}  +\frac{1}{2}  |\vec v|^2) \partial_t \tilde \rho + \tilde \rho \vec v \partial_t \vec v \right ),
\]
where the chemical potential $\tilde \mu_{\rm c} = \delta F_{\tilde \rho}/\delta \tilde \rho$ is now a function of both $\tilde \rho$ and its gradients.
In a one-component homogeneous liquid, as formally discussed in Ref. \cite{martinUnifiedHydrodynamicTheory1972}, mass transport occurs through advection by the hydrodynamic velocity field only.
In crystals, for examples, microscopic dissipative currents may be present and mediate dislocation motion and vacancy diffusion.
Thus, we rewrite Eq.~(\ref{eq:liquid_conservation_of_mass}) by including a microscopic current $\tilde {\vec J}$ as a source term
\[
\partial_t \tilde \rho + \nabla \cdot (\tilde \rho \vec v)= \nabla \cdot \tilde \vec {J}.
\label{eq:microscopic_structure_continuity_equation}
\]
$\tilde {\vec J}$ will be determined by free energy minimization and should reduce to ($\tilde {\vec J}=\vec 0$) for homogeneous liquid systems.
To connect the microscopic degrees of freedom ($\tilde \rho,\nabla \tilde \rho,...$) with the (coarse) hydrodynamic variables ($\vec v,\sigma,...$), we need to formalize the notion of spatial coarse-graining, which is here defined through the Gaussian convolution
\[
\langle {\tilde X} \rangle (\vec r) = \int d\vec r'\frac{\tilde X(\vec r')}{(2\pi w^2)^{d/2}} \exp\left (-\frac{(\vec r-\vec r')^2}{2w^2}\right),
\label{eq:coarse_graining_operation}
\]
where $d$ is the dimension of the system and $\langle \tilde X \rangle$ is the slowly-varying component of $\tilde X$.
The velocity field $\vec v$ remains a field that is slowly-varying over the microscopic length scale $w$ and thus satisfies the property that
\[\langle \vec v \tilde X \rangle = \vec v\langle \tilde X \rangle.
\label{eq:slowly_varying_property}
\]
Inserting Eq.~(\ref{eq:microscopic_structure_continuity_equation}) into the  momentum balance, Eq.~(\ref{eq:momentum}), and applying the coarse-graining procedure, we find that
\[
\langle \tilde \rho \rangle D_t  \vec v = \nabla \cdot  \sigma   -\vec v \langle \nabla \cdot \tilde{\vec J}\rangle.
\label{eq:momentum_equation_coarse_grained}
\]
Thus, the evolution of the free energy reduces to
\[
\begin{array}{ll}
\partial_t F &= \int d \vec r\left ( (\tilde \mu_{\rm c} +\frac{1}{2}  |\vec v|^2)  (\nabla \cdot \tilde{\vec J} - \nabla \cdot (\tilde \rho \vec v)) \right .\\
& \left. ~~~~~~~~~~~~~
+ \vec v \cdot (\nabla \cdot  \sigma - \rho (\vec v \cdot \nabla) \vec v -\vec v \nabla \cdot \tilde{\vec J}) \right ).
\end{array}
\]
Again, by repeated integration by parts and using that $\tilde \rho \nabla \tilde \mu_{\rm c} = -\nabla \cdot \tilde \sigma^{(\tilde \rho)}$ where $\tilde \sigma^{(\tilde \rho)}$ is the microscopic reversible stress tensor for a density gradient based free energy  \cite{skogvollStressOrderedSystems2021}, we get
\[
\partial_t F
= \int d\vec r\left [ -\tilde {\vec J}\cdot  \nabla (\tilde \mu_{\rm c} -\frac{1}{2}  |\vec v|^2)     -  ( \sigma  -  \tilde{\sigma}^{(\tilde \rho)}) : (\nabla \vec v)  \right ].
\]
In the absence of dissipation and external forces, $\tilde {\vec J}=\vec 0$, the stress is given by $\sigma =\tilde{\sigma}^{(\tilde \rho)}$.
At first glance, this might look like an inconsistency, since we are equating a quantity that is slowly varying with one featuring rapid variations at the atomic length scale.
However, as evidenced in Eq.~(\ref{eq:momentum_equation_coarse_grained}), it is the coarse-grained stress that couples with the velocity field.
This is reflected in the evolution of the free energy by noting that coarse-graining the integrand does not affect the total free energy, i.e., $\int d\vec r \tilde X = \int d\vec r \langle \tilde X \rangle$.
Thus, by coarse-graining the second term in the integrand and using the property of slowly-varying variables, Eq.~(\ref{eq:slowly_varying_property}), we get
\[
\partial_t F
= \int d\vec r\left [ -\tilde {\vec J}\cdot  \nabla (\tilde \mu_{\rm c} -\frac{1}{2}  |\vec v|^2)     -  (   \sigma  -  \langle \tilde{\sigma}^{(\tilde \rho)} \rangle) : (\nabla   \vec v)  \right ],
\]
showing that the stress that couple the velocity field in a self-consistent way is the coarse-grained reversible stress $\langle \tilde{\sigma}^{(\tilde \rho)} \rangle$, which provides a more rigorous argument for the introduction of coarse-graining in Ref. \cite{skogvollStressOrderedSystems2021}.

Using the same linear response arguments as in the previous section, the dissipative stress takes the same form, i.e., $  \sigma -  \sigma^{(\rm{ext})} -  \langle \tilde{\sigma}^{(\tilde \rho)} \rangle = \mathcal D : \nabla   \vec v$.
Similarly, the microscopic dissipative current $\tilde {\vec J}$, as the conjugate of $\tilde \mu_{\rm c}-\frac{1}{2}  |\vec v|^2$, is constrained to be $\tilde {\vec J} = \mathcal Q \cdot \nabla (\tilde \mu_{\rm c} -\frac{1}{2}  |\vec v|^2)$, where $\mathcal Q$ is a rank-two mobility tensor.
As previously alluded to, we expect no dissipative mass flow ($\tilde {\vec J}=\vec 0$) in the (single-component) liquid phase.
Thus, the mobility tensor $\mathcal Q$, in addition to being positive definite, should be informed about the microscopic structure in such a way that it is identically zero in the liquid phase.
To this end, we provide the following heuristic argument.
Consider a microscopically varying function, given by a profile $\tilde \rho(x) = \eta e^{iqx} + \eta^* e^{-iq x}$, for a complex amplitude $\eta$, such that the variations vanish upon coarse-graining, i.e., $\langle e^{iqx}\rangle=0$. The mode of microscopic diffusion transport should scale with both the amplitude $\eta$ (whose nonzero value denotes such variations) and the size of $q$ (larger $q$-values give shorter spatial variations enabling more diffusion). A candidate for $\mathcal Q\ge 0$ satisfying these properties is $\mathcal Q = \Gamma \mathcal S^{(\tilde \rho)} \equiv \Gamma \langle (\partial_x \tilde \rho) (\partial_x \tilde \rho) \rangle$, where $\Gamma>0$ is a constant parameter.
To see this, we note that $\langle (\partial_x \tilde \rho) (\partial_x \tilde \rho) \rangle = \langle (iq \eta e^{iqx} -iq \eta^* e^{-iq x})(iq\eta e^{iqx} - iq\eta^* e^{-iq x})\rangle = 2|\eta|^2 q^2$.
Generalizing this argument to higher dimensions, we thus postulate a mobility coefficient tensor constructed from the density gradients as
\[
\mathcal Q = \Gamma \mathcal S^{(\tilde \rho)} = \Gamma \langle (\nabla \tilde \rho)(\nabla \tilde \rho) \rangle.
\]
$\mathcal S^{(\tilde \rho)}$ is a symmetric, rank-two tensor called the structure tensor and satisfies the requirement of positive definiteness by definition.
This tensor has wide application for tracking features in patterns and microscopic structures as it allows for characterizing fields locally together with the distribution of their gradients and, in turn, direction-dependent properties (see, e.g., Refs. \cite{asipauskas2003texture,laptev2005space,akl2015texture,mueller2019,Wenzel2021}).
In particular, as a symmetric tensor, in two dimensions it has two orthogonal eigenvectors $(\vec e_1, \vec e_2)$ with real eigenvalues $(\lambda_1,\lambda_2)$. The microscopic density current is then given by
\[
\tilde {\vec J} =
 \lambda_1 \Gamma (\partial_1 \tilde \mu_{\rm c}^{\rm eff}) \vec e_1
+ \lambda_2 \Gamma (\partial_2 \tilde \mu_{\rm c}^{\rm eff})\vec e_2,
\label{eq:diffusion_along_eigenvectors}
\]
where $\partial_i = \vec e_i \cdot \nabla$, the directional derivative along $\vec e_i$ and $\tilde \mu_{\rm c}^{\rm eff} = \tilde \mu_{\rm c} -\frac{1}{2}  |\vec v|^2$.
Thus, the mass diffusion only takes place along a direction if the structure tensor $\mathcal S^{(\tilde \rho)}$ has a non-zero eigenvalue, and the effective chemical potential has a gradient in that direction.
In the case of crystal density profiles, it reduces to an isotropic tensor $\mathcal S^{(\tilde \rho)}_{ij} \propto \delta_{ij}$, while in the liquid phase, it is identically zero.
Figures~\ref{fig:MicroStructures}(d-f) shows the structure tensor $\mathcal S^{(\tilde \rho)}$ for the patterns/microscopic structures in Figures~\ref{fig:MicroStructures}(a-c), respectively.
This leads to the full set of equations
\[
\left \{
\begin{array}{ll}
\partial_t \tilde \rho = \Gamma \nabla \cdot [\mathcal S^{(\tilde \rho)} \cdot \nabla (\tilde \mu_{\rm c} -\frac{1}{2}  |\vec v|^2)] - \nabla \cdot (\tilde \rho \vec v),  \\
\langle \tilde \rho \rangle D_t \vec v = -\langle \tilde \rho \nabla \tilde \mu_{\rm c} \rangle +  \vec f^{(\rm{ext})}  + \nabla\cdot \left[\mathcal D : \nabla   \vec v\right].
\end{array}
\right .
\label{eq:fullHPFC}
\]

\paragraph{Liquid limit:} This corresponds to $\mathcal S^{(\tilde \rho)} = 0$, in which case $\tilde {\vec J} = 0$, and the model reduces to that of Section \ref{sec:classical_hydrodynamics}. We thus recover the classical hydrodynamic equations of homogeneous and isotropic fluids.

\paragraph{Crystal phase:}

Suppose that the free energy density $\tilde f(\tilde \rho,\nabla \tilde \rho,...)$ is such that the minimizer function $\tilde \rho^{\rm eq}$ of the free energy has a crystalline structure, given by some reciprocal lattice vector structure $\mathcal R$, i.e., $\tilde \rho^{\rm eq} = \sum_{\vec q \in \mathcal R} \eta_{\vec q} {\rm e}^{i\vec q\cdot \vec r}$ with $i$ the imaginary unit.
In Ref. \cite{skogvollPhaseFieldCrystal2022}, we have shown that for a large class of symmetric lattices, we get $\mathcal S^{(\tilde \rho^{\rm eq})}_{ij} = \kappa \delta_{ij}$, where $\kappa$ is a proportionality constant depending on the structure of $\tilde \rho$, which leads to isotropic diffusion in addition to the advection by the macroscopic velocity field $\vec v$. 
In the case of a crystal order parameter, it was shown in Ref. \cite{skogvollStressOrderedSystems2021} that the force density $\nabla \cdot \langle \tilde \sigma^{(\tilde \rho)} \rangle = -\langle \tilde \rho \nabla \tilde \mu_{\rm c} \rangle$ can be decomposed into the divergence of an isotropic pressure $-\nabla P = \nabla (\langle \tilde f \rangle -\langle \tilde  \mu_{\rm c}\tilde \rho\rangle )$ and the divergence of a rank-two tensor $h^{(\tilde \rho)}$, which is interpreted as the thermodynamic conjugate of the strain, and has a divergence given by $\nabla \cdot h^{(\tilde \rho)} = \langle \mu_{\rm c} \nabla \tilde \rho - \nabla \tilde f\rangle$.
The momentum equation then reduces to
\[
\langle \tilde \rho \rangle D_t \vec v = -\nabla P + \nabla \cdot h^{(\tilde \rho)} +  \vec f^{(\rm{ext})}  + \nabla \cdot [ \mathcal D : \nabla   \vec v].
\]

\section{Phase field crystal modeling}

In the rest of this paper, we consider the phase field crystal (Swift-Hohenberg) free energy in terms of the rapidly varying density field, denoted $\tilde\rho \equiv \psi$.
It can be written as
\begin{equation}
F_\psi=\int_{\Omega} \left[\frac{\Delta \texttt B_0}{2}\psi^2+\frac{\texttt B^x_0}{2} \psi\mathcal L^2\psi
-\frac{\texttt t}{3}\psi^3+\frac{\texttt v}{4}\psi^4 \right]d\mathbf{r},
\label{eq:F_PFC}
\end{equation}
with $\Omega$ the integration domain, and $(\Delta \texttt B_0, \texttt B^x_0, \texttt t,\texttt v)$ parameters that may be fitted to material properties \cite{elderPhasefieldCrystalModeling2007}.
$\mathcal L$ is a derivative operator which will take two different forms: $\mathcal L_{1M} = (q_0^2+\nabla^2)$ for systems with 2D hexagonal or 3D bcc lattice symmetries and $\mathcal L_{2M} = (q_0^2 + \nabla^2)(2q_0^2 + \nabla^2)$ for the 2D square lattice \cite{mkhontaExploringComplexWorld2013,emdadiRevisitingPhaseDiagrams2016}.
The parameter $q_0$ gives the length scale $w= 2\pi q_0^{-1}$ of the forming microscopic structures which will be used for the coarse-graining operation, Eq.~(\ref{eq:coarse_graining_operation}).
The minimizer of this free energy for a certain range of parameters will be a spatially periodic function $\psi^{\rm eq}(\vec r)$ of a given microscopic structure.
The corresponding chemical potential for this free energy is given by
\[
\tilde \mu_{\rm c} = \frac{\delta F_\psi}{\delta \psi} = ( \Delta \texttt B_0  + \mathcal L^2 )\psi - \texttt t \psi^2 + \texttt v \psi^3.
\]
Usually, the phase field free energy is supplied with a conservative evolution equation which reduces its free energy
\[
\partial_t \psi = \Gamma \nabla^2 \tilde \mu_{\rm c}.
\qquad
\textrm{(The PFC model)}
\label{eq:PFC_dynamics}
\]
We will refer to Eq.~(\ref{eq:PFC_dynamics}) as \textit{\textit}{the PFC model}.
Given a crystal equilibrium state $\psi^{\rm eq}$, the lattice constant $a_0 \sim w$ will serve as a characteristic unit of length and the shear modulus $\mu$ \cite{skogvollStressOrderedSystems2021} as a characteristic unit of stress.
Eq.~(\ref{eq:PFC_dynamics}) is appropriately rescaled using $\tau = (\Gamma \texttt B_0^x q_0^6)^{-1}$ as a characteristic unit of time.

The full set of equations, Eq.~\ref{eq:fullHPFC}, is a set of highly non-linear differential equations. We consider two limits where simplifying assumptions can be considered: negligible wave dynamics, i.e. $\mathbf{v}=0$, and taking a  constant average density $\langle \tilde \rho \rangle=\rho_0$.

\subsection{Zero wave dynamics}

\begin{figure}[htp]
    \centering
    \includegraphics[width=\textwidth]{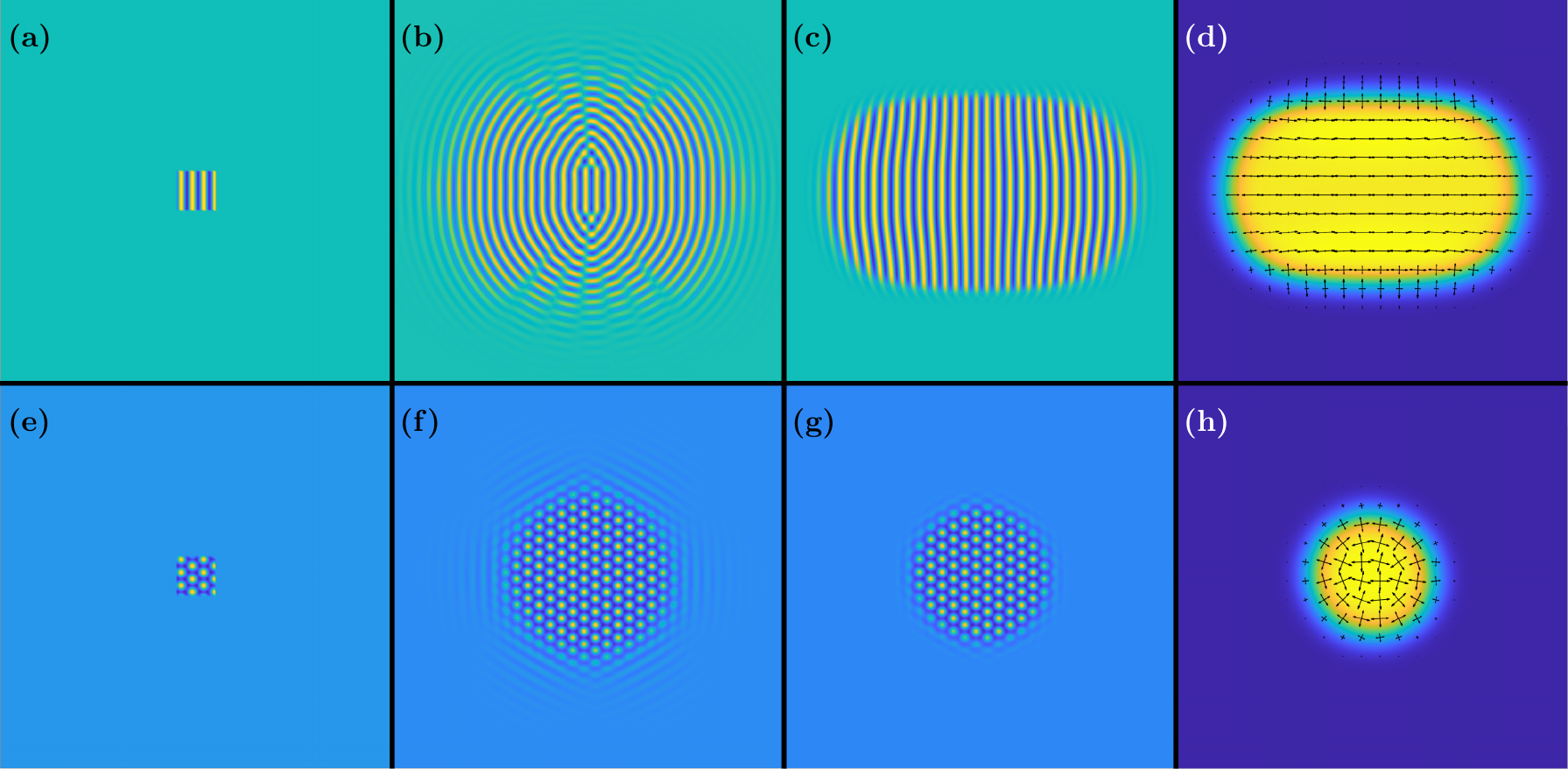}
    \caption{Growth of a (a) stripe and (e) triangular phase in a super-cooled liquid phase (see text for simulation details).
    The second column shows the fields after evolving according to PFC dynamics, Eq.~(\ref{eq:PFC_dynamics}), at (b) $t=25\tau$ and (f) $t=80\tau$.
    The third column shows the fields after evolving according to the model with the spatially dependent mobility, Eq.~(\ref{eq:spatial_dependent_diffusion}), at (c) $t=250\tau$ and (g) $t=400\tau$, in which diffusion only happens in the directions of the  eigenvectors of the structure tensor $\mathcal S^{(\psi)}$, shown in (d) and (h), respectively, in the same presentation as in Fig.~\ref{fig:MicroStructures}.
    In (d) and (h), the colorbars range between $[0,2q_0^2\eta_{0,st}^2]$ and $[0, 3 \sqrt{2} q_0^2\eta_{0,tr}^2]$, respectively, corresponding to $\sqrt{\lambda_1^2 + \lambda_2^2}$ for the equilibrium structure tensors in Eq.~(\ref{eq:equlibrium_Structure_tensors}).}
    \label{fig:comparison_spatially_dep_diffusion}
\end{figure}

For a negligible velocity field ($\mathbf{v}=\vec 0$), the model resembles the classical PFC model. However, the major difference is that we have a spatially dependent mobility coefficient, namely
\[
\partial_t \psi = \Gamma \nabla\cdot\left[ \mathcal S^{(\psi)} \cdot \nabla \tilde \mu_{\rm c}\right],
\label{eq:spatial_dependent_diffusion}
\]
which is also a highly nonlinear contribution to the evolution equation. This contribution is specifically important at the interface between two different phases.
To highlight the effect of the spatially dependent mobility in setting the interfacial properties, we consider the evolution of an interface during a symmetry-breaking phase transition, namely crystallization.
This is achieved by initializing the phase field $\psi$ in the under-cooled liquid (disordered) phase $\psi(\vec r)=\psi_0$ for a set of parameters where the equilibrium is an ordered state, e.g., stripes or crystal lattices.
We use the set of parameters $(q_0,\Delta \texttt B_0,\texttt B^x_0,\texttt t,\texttt v)=(1,-0.3,1,0,1)$ and $\mathcal L = \mathcal L_{1M}$ for the free energy, Eq.~(\ref{eq:F_PFC}).
Since the dynamics of Eq.~(\ref{eq:spatial_dependent_diffusion}) is conservative, the average density $\psi_0$ is also a parameter. For $\psi_0=0$, the minimizer $\psi^{\rm eq}(\vec r)$ is a stripe phase $\psi_{\rm st} \approx \psi_0 + \eta_{0,{\rm st}} \sum_{\{\vec q\}} {\rm e}^{i\vec q\cdot \vec r} + {\rm c.c.}$ for $\{ \vec q\} = \{ (1,0)\}$ and for $\psi_0=-0.3$, a triangular phase $\psi_{\rm tr} \approx \psi_0 + \eta_{0,{\rm tr}} \sum_{\{\vec q\} } {\rm e}^{i\vec q\cdot \vec r} + {\rm c.c.}$ for $\{ \vec q \} = \{(0,1),(\frac{\sqrt 3}{2},-\frac{1}{2}),(-\frac{\sqrt 3}{2},-\frac{1}{2})\}$ \cite{Elder2004}.
The equilibrium amplitude $\eta_0$ can in both cases be determined by inserting the ansatzes into Eq.~(\ref{eq:F_PFC}) and minimizing with respect to $\eta_0$.
For this set of parameters, we get $\eta_{0,st} = 0.1368$ and $\eta_{0,tr} = 0.3162$ for the stripe and triangular state, respectively.
The equilibrium structure tensors are given by
\[
\mathcal S^{(\psi_{\rm st})} =  2q_0^2 \eta_{0,{\rm st}}^2
\left (
\begin{array}{cc}
1 & 0 \\
0 & 0 \\
\end{array}
\right ) ,
\qquad \mathcal S^{(\psi_{\rm tr})} = 3 q_0^2 \eta_{0,{\rm tr}}^2
\left (
\begin{array}{cc}
1 & 0 \\
0 & 1 \\
\end{array}
\right ),
\label{eq:equlibrium_Structure_tensors}
\]
so diffusion takes place only along the direction perpendicular to the stripes for the stripe phase (see Fig.~\ref{fig:MicroStructures}(c),(f)) and it is isotropic for the triangular phase.
In order to compare the time scale of results from the simulations with the spatially dependent mobility from Eq.~(\ref{eq:spatial_dependent_diffusion}) and the classical PFC model with constant mobility in Eq.~(\ref{eq:PFC_dynamics}), we scale the dissipation parameter $\Gamma$ in the former equation by $\Gamma_{\rm st} = (2q_0^2 \eta_{0,{\rm st}}^2)^{-1} \Gamma$ and $\Gamma_{\rm tr}=(3q_0^2 \eta_{0,{\rm tr}}^2)^{-1} \Gamma$ for the simulations with the striped and triangular samples, respectively.
The parameter $\Gamma$ was set to $1$ in these simulations.
For both cases, we prepare a $20w\times20w$ periodic computational domain, with $w=2\pi q_0^{-1}$, and set $\psi(\vec r)=\psi_0$ everywhere except for a $2w\times 2w$ region in the middle hosting the corresponding equilibrium configuration.
A Fourier pseudo-spectral method is employed for spatial discretization while the time evolution for the two models, i.e. Eq.~(\ref{eq:PFC_dynamics}) and Eq.~(\ref{eq:spatial_dependent_diffusion}), was computed by an exponential time differencing method \cite{coxExponentialTimeDifferencing2002} with $\Delta t=0.1$ and a forward Euler scheme with $\Delta t=10^{-4}$, respectively, until the forming pattern had neared the boundary of the simulation domain.
Note that the numerical integration of the model featuring the spatially dependent mobility requires a very small time step  in the considered explicit scheme.
This is adopted here to provide a first illustration of the model, while improvements will be explored in future works for larger-scale investigations.
The result of numerical simulations is shown in Fig.~\ref{fig:comparison_spatially_dep_diffusion}, and several qualitative differences between the two types of diffusion are observed.
With the spatially dependent mobility, the interface is sharper, thus more localized, and has slower dynamics.
This is expected as the diffusion only takes place at the non-zero value of the structure tensor, which scales with the amplitude. Therefore, the microscopic structure grows essentially only at the boundary of the fully formed structure.
This effect is as intended since the model is shown to suppress diffusion behavior in the liquid phase.
Due to the isotropic mobility in the PFC evolution of the striped phase, panel (b), interference patterns from the growth on each side of the seed appear, which are not present in the case with the spatially dependent mobility.
In the latter case, the anisotropy of the initial seed causes the seed to grow faster in the direction perpendicular to its structure, as expected from the form of the structure tensor, as shown in panel (d).
For the simulation parameters chosen here, the growth of the stripe phase is $\approx 50\%$ faster in the direction perpendicular to the stripes. The anisotropy in the growth rate may depend on model parameters and an extensive study is beyond the scope of
this paper.
The study indicates, however, that during growth processes, the type of mobility used in the modeling is important for the resulting macroscopic structure.

\subsection{Constant average density}

Having examined how the model behaves in the zero-wave dynamics limit in terms of evolving interfaces, we now shift focus for the rest of the paper to study the coupling of the dissipative PFC dynamics with the wave dynamics and apply this framework to the dynamics of bulk crystals with a constant density $\rho_0$.
In this regime, Eq.~(\ref{eq:PFC_dynamics}) reduces to isotropic diffusion coupled with a momentum equation,
\[
\left \{
\begin{array}{l}
\partial_t \psi = \Gamma \nabla^2 \tilde \mu_{\rm c} - \vec v \cdot \nabla \psi
\\
\rho_0 \partial_t \vec v = \langle \tilde \mu_{\rm c} \nabla \psi - \nabla \tilde f  \rangle + \Gamma_S \nabla^2 \vec v + \vec f^{(\rm{ext})}.
\end{array} \right . \quad \textrm{(The sHPFC model)}
\label{eq:sHPFC_dynamics}
\]
For simplicity, we have not included the isotropic pressure part of the stress as this part is primarily due to variations in the average density.
A state of mechanical equilibrium is thus characterized by $\nabla \cdot h^{(\psi)} =\langle \tilde \mu_{\rm c} \nabla \psi - \nabla \tilde f  \rangle = 0$.
Being a simplified hydrodynamic framework, similar to that of Ref.\cite{podmaniczkyNucleationPostNucleationGrowth2021}, we use the acronym sHPFC in this paper.
In Ref. \cite{skogvollPhaseFieldCrystal2022}, the PFCMEq model was introduced to evolve the PFC model at mechanical equilibrium as
\[
\left \{
\begin{array}{l}
\partial_t \psi = \Gamma \nabla^2 \tilde \mu_{\rm c} \\
\psi(\vec r) \rightarrow \psi(\vec r-\vec u^{\delta}) \textrm{ every }\Delta t
\end{array}
\right .
\qquad \textrm{(The PFCMEq model)}
\label{eq:PFCMEq_dynamics}
\]
where $\vec u^\delta$ is a displacement encoding the deformation of $\psi$ to be performed at every numerical time step $\Delta t$ to maintain mechanical equilibrium.
$\vec u^\delta$ is found numerically by solving the elastic problem in the medium and requires the knowledge of the elastic constants that are derived from equilibrium properties.
Although successful in evolving at mechanical equilibrium, the PFCMEq has some shortcomings.
Indeed, the requirement on the \textit{a priori} knowledge of elastic constants, which are well defined only for relaxed bulk crystals with known orientation, poses issues for the investigation of crystal/liquid interfaces, large strains, and anisotropic polycrystals.
We use this model as a benchmark for the sHPFC model, which is then shown to overcome most of these limitations.

\section{Properties and characterization of the sHPFC model}

\subsection{Small deformation limit}

The Swift-Hohenberg free energy is postulated on the grounds that it encodes all equilibrium properties of the different ordered phases, and this includes thus also the elastic constants determined by the crystal lattice symmetries. However, as discussed in the introduction, the diffusive evolution of the phase field cannot capture alone any elastic phenomena.
The main motivation of a hydrodynamic extension of the classical PFC is precisely to couple the diffusive PFC dynamics with wave dynamics on scales much larger than the lattice unit.
To this end, we show that the momentum equation in the sHPFC model reduces to the wave equation in the small-deformation limit. We consider a small perturbation $\delta \psi$ of the equilibrium PFC density of a perfect crystal
\[\psi = \psi^{\rm eq}+\delta\psi, \]
where $\psi^{\rm eq}$ is the equilibrium crystal given in the one-mode approximation as
\[
\psi^{\rm eq} = \psi_0 + \eta_0 \sum_{n=1}^N e^{i\vec q^{(n)}\cdot \vec r}\]
where $\{ \vec q^{(n)} \}_{n=1}^N$ are the $N$ smallest reciprocal lattice vectors (i.e., with length $|\vec q^{(n)}|=q_0$)
and the perturbation is due to a small deformation $\vec u(\vec r,t)$, so that
\[
\delta\psi = - \vec u\cdot \nabla \psi^{\rm eq}.\]
Linearizing the evolution equations around an equilibrium state, i.e. $\psi=\psi^{\rm eq}$ and $\vec v=\vec 0$, we get
\[
-\nabla \psi^{\rm eq} \cdot \partial_t \vec u = \Gamma \nabla^2 \delta \tilde \mu_{\rm c} - \delta \vec v \cdot \nabla \psi^{\rm eq},
\label{eq:dtu_linear_stab_pre_elim}
\]
\[
\rho_0 \partial_t \delta \vec v = \nabla \cdot (\mathcal C :  \nabla   \vec u)  + \Gamma_S \nabla^2 \delta \vec v,
\]
where we used that $\partial_t\delta\psi = -\partial_t \vec u\cdot \nabla \psi^{\rm eq}$,
and $\mathcal C$ is the rank-four elastic constants tensor at constant PFC density, which depends on the crystal symmetry (for more details, see Ref. \cite{skogvollStressOrderedSystems2021}).
To obtain a term dependent on $\vec u$ only, we multiply Eq.~(\ref{eq:dtu_linear_stab_pre_elim}) with $\nabla \psi^{\rm eq}$ and apply coarse-graining.
Using the property of the coarse-grained quantities, Eq.~(\ref{eq:slowly_varying_property}), and that for a large class of symmetric lattices $\langle \partial_i \psi^{\rm eq} \partial_j \psi^{\rm eq} \rangle =  q_0^2\eta_0^2 \frac{N}{d} \delta_{ij}$, where $d$ is the dimension of the system \cite{skogvollPhaseFieldCrystal2022}, we get
\[
q_0^2 \eta_0^2\frac{N}{d} \partial_t \vec u = -\Gamma \langle (\nabla \psi^{\rm eq} )\nabla^2 \delta \tilde \mu_{\rm c} \rangle + q_0^2 \eta_0^2 \frac{N}{d} \delta \vec v.
\]
Inside the coarse-graining, we can apply partial integration (move derivatives).
Therefore, using that in the one mode approximation, $\nabla^2 \psi = - q_0^2 \psi^{\rm eq} + q_0^2 \psi_0$, we can rewrite the first term as follows
\[
-\Gamma \langle (\nabla \nabla^2 \psi^{\rm eq} )\delta \tilde \mu_{\rm c} \rangle
= \Gamma \delta \langle \tilde \mu_{\rm c} \nabla \psi^{\rm eq} - \nabla \tilde f \rangle
= \Gamma \nabla \cdot (\mathcal C : \nabla   \vec u),
\]
where we have used that $\delta \langle \tilde f \rangle  = 0$ and that differentiation commutes with coarse-graining.
This gives
\[
\partial_t \vec u = \delta \vec v +\frac{\Gamma d}{N q_0^2 \eta_0^2} \nabla \cdot (\mathcal C : \nabla   \vec u).
\]
Combining the two previous equations, we get a damped, wave equation for the displacement field, with $\Gamma_S=0$
\[
\partial_t^2 \vec  u = \frac{1}{\rho_0} \nabla \cdot (\mathcal C : \nabla   \vec u) + \frac{d}{Nq_0^2\eta_0^2} \partial_t (\nabla \cdot (\mathcal C : \nabla   \vec u)).
\]
Inserting the isotropic elastic constants $\mathcal C_{ijkl} = \lambda \delta_{ij}\delta_{kl} + 2\mu \delta_{k(i} \delta_{j)l}$, we get
\[
\partial_t^2 \vec  u =\frac{1}{\rho_0} (\mu \nabla^2 \vec u + (\lambda+\mu)\nabla \nabla \cdot \vec u)
+ \frac{\Gamma d}{Nq_0^2\eta_0^2} \partial_t (\mu \nabla^2 \vec u + (\lambda+\mu)\nabla \nabla \cdot \vec u),
\]
and, furthermore, for the hexagonal 2D lattice with $q_0=1$, where $N=6$, $d=2$ and $\mu=\lambda=3\texttt B^x\eta_0^2$, we find
\[
\partial_t^2 \vec  u =3 \texttt B^x \eta_0^2\rho_0^{-1} ( \nabla^2 \vec u + 2\nabla \nabla \cdot \vec u)
+ \Gamma \texttt B^x \partial_t (\nabla^2 \vec u + 2 \nabla \nabla \cdot \vec u).
\label{eq:PFC_hex_wave_equation}
\]
In the limit of vanishing microscopic dissipation ($\Gamma=0$), this reduces to a classical wave equation of a perfectly elastic medium.
For $\Gamma\neq 0$, we see that the classical evolution of the PFC introduces a mode of momentum dissipation.
In Ref. \cite{heinonenConsistentHydrodynamicsPhase2016}, the same wave equation Eq.~(\ref{eq:PFC_hex_wave_equation}) was obtained from the amplitude formulation, and shown that it has a dispersion relation which is linear in the long-wavelength limit.

Our derivation of Eq.~(\ref{eq:PFC_hex_wave_equation}) is from a microscopic density field instead of slowly-varying amplitudes. Therefore, the current approach is more versatile with a larger range of possible applications.  To illustrate this point, we consider next a fundamental problem in dislocation dynamics, namely dislocation annihilation, and highlight a couple of other applications of the model.

\subsection{Dislocation annihilation}
In Ref. \cite{skaugenDislocationDynamicsCrystal2018}, it was found that dislocations in the 2D triangular PFC model have overdamped dynamics driven by the Peach-Koehler force and a mobility $M=(4\pi\eta_0^2)^{-1}$.
In an isotropic and homogeneous elastic medium, dislocations separated by a distance $d$ move with a velocity
\[
v = K_T d^{-1},
\label{eq:Annihilation_theoretical_result}
\]
where $K_T = 2Mb \mu/(3\pi)$ and is given in units of $a_0^2/\tau$ \cite{skaugenDislocationDynamicsCrystal2018}.

\begin{figure}
    \centering
    \includegraphics[width=\textwidth]{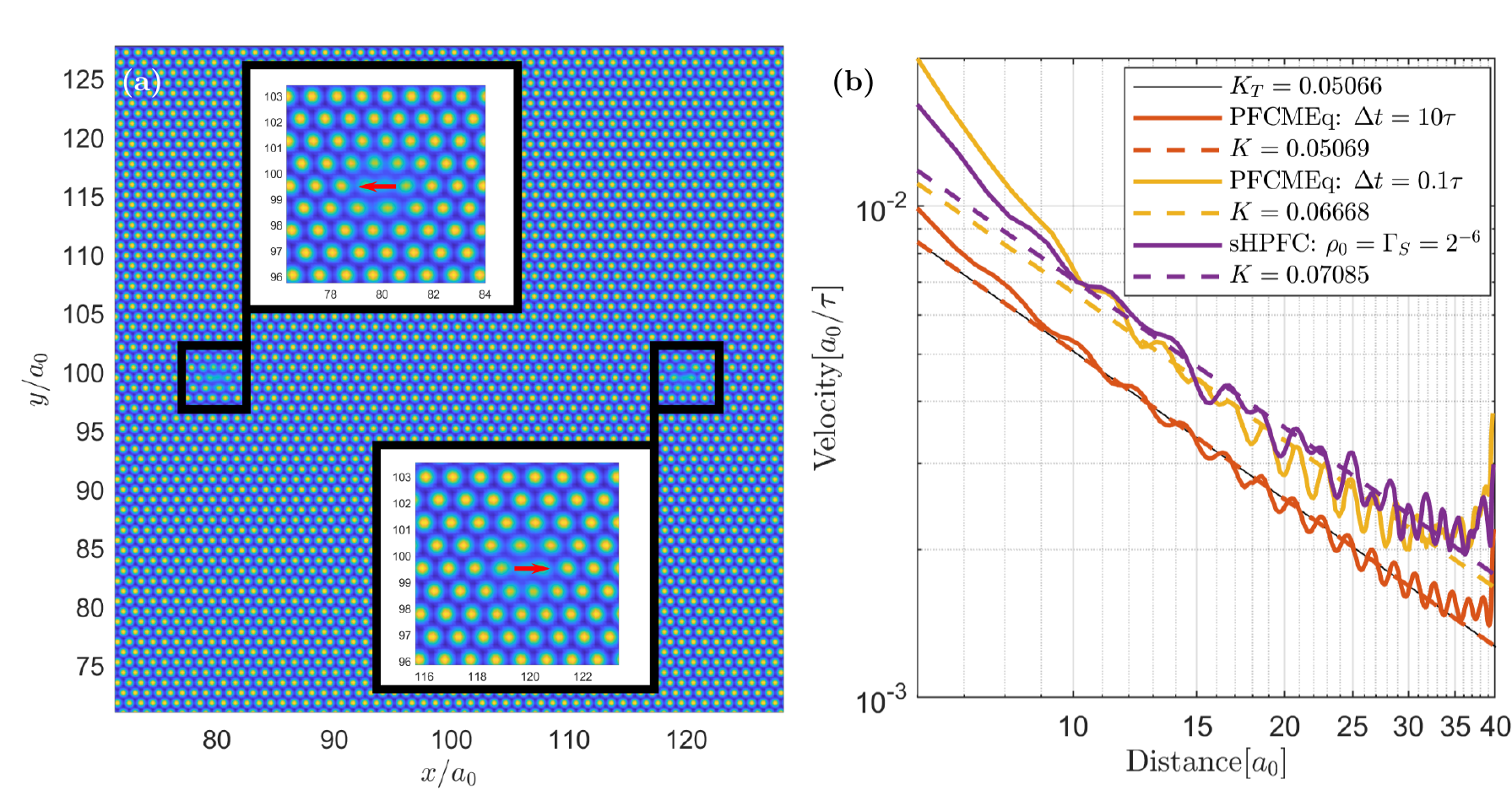}
    \caption{Annihilation by gliding of two dislocations in a two-dimensional crystal with triangular symmetry. (a)
    The initial PFC configuration seeded with (inset) the dislocation dipole with Burgers vectors $\pm a_0 \vec e_x$ (red quivers).
    (b) The velocity of the dislocations as a function of their distance.
    The black line shows the theoretical result of Eq.~(\ref{eq:Annihilation_theoretical_result}).}
    \label{fig:benchmark_setup_and_PFCMEq}
\end{figure}

In this section, we benchmark the sHPFC model against this classical prediction of linear elasticity theory.
We prepared a $200\times200$ unit cell triangular PFC configuration with a dislocation dipole (Burgers vectors $\vec b = \pm a_0 \vec e_x$) at $x=80a_0$ and $x=120a_0$.
Parameters used were $(q_0,\Delta \texttt B_0,\texttt B^x_0,\texttt t,\texttt v,\psi_0)=(1,-0.2,1,0,1,-0.265)$ and $\mathcal L = \mathcal L_{1M}$ to match those of Ref. \cite{skaugenDislocationDynamicsCrystal2018}.
With these parameters, we get the following theoretical value of $K_T = 0.05066 a_0^2/\tau$.
We observed that the annihilation in all cases qualitatively follow the same trajectory
\[v=Kd^{-1},\]
and so for each simulation we extract an estimate of mobility factor $K$ by linear regression.
Similar to the simulations reported in the previous section, the Fourier pseudo-spectral method was employed for spatial discretization.
The velocity field $\vec v$ was evolved by a forward Euler scheme, the phase field advection term ($-\vec v \cdot \nabla \psi$) by a midpoint Euler scheme, and the gradient descent contribution ($\nabla^2 \tilde \mu_{\rm c}$) by an exponential time differencing method \cite{coxExponentialTimeDifferencing2002}.
A temporal resolution of $\Delta t=0.1$ was used.
Figure \ref{fig:benchmark_setup_and_PFCMEq} shows the initial setup and the dislocation velocity as a function of distance.
The sHPFC model was used with $\rho_0 = \Gamma_S = 2^{-6}$ and compared to the results of the PFCMEq model, which has previously been shown to give a correct description of dislocation interaction.
Since the latter method imposes a compatible distortion at every time step, the time interval $\Delta t$ sets the effective speed of relaxation to elastic equilibrium \cite{skogvollPhaseFieldCrystal2022}.
This leads to a velocity increase for smaller time step and we see that the sHPFC model with the given parameters closely matches the kinematic path of the PFCMEq with $\Delta t = 0.1\tau$.

\begin{figure}
    \centering
    \includegraphics[]{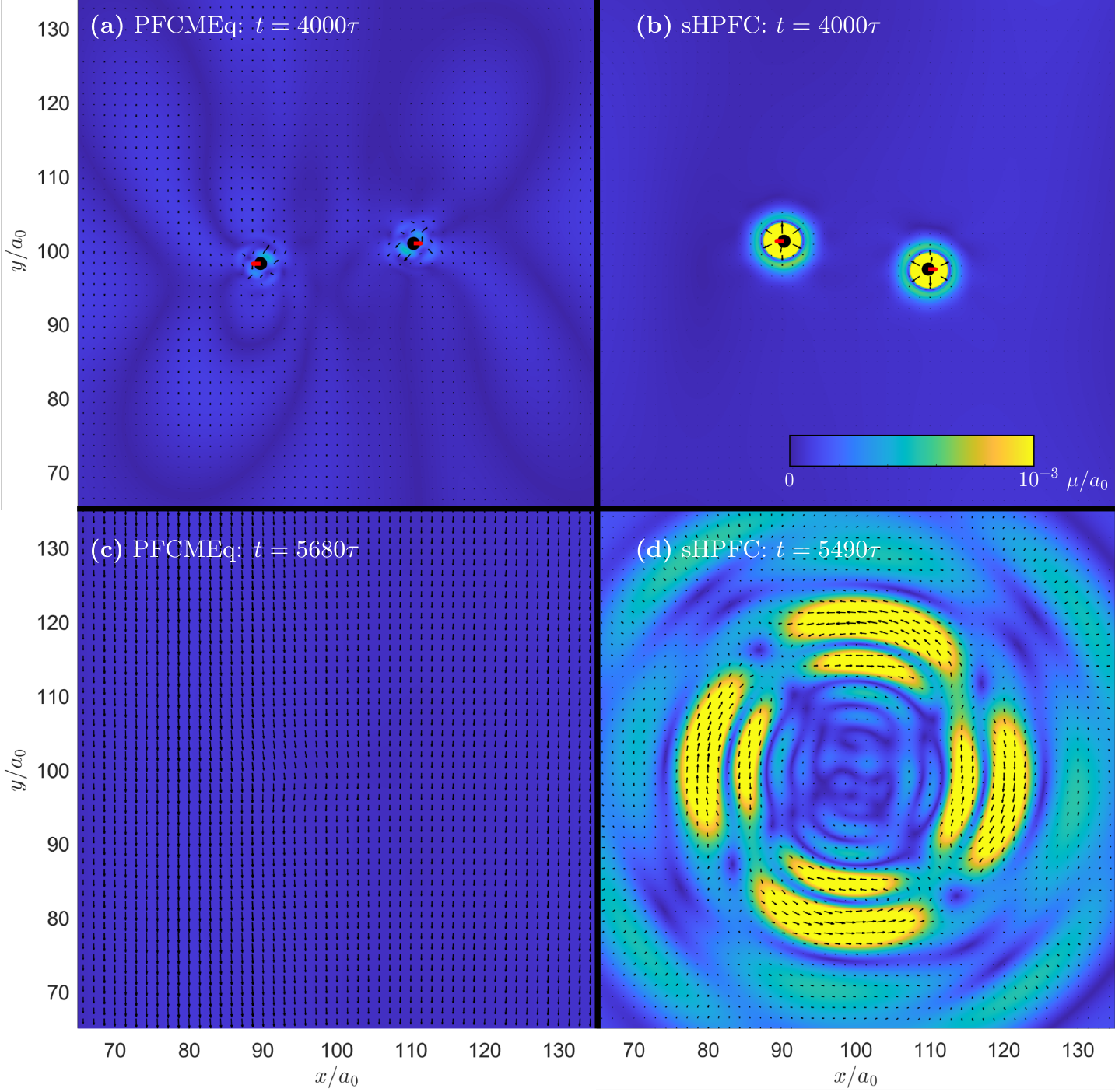}
    \caption{The force density $\nabla \cdot h^{(\psi)} = \langle \tilde \mu_{\rm c} \nabla \psi - \nabla \tilde f \rangle$ during the annihilation process for the different models.
    The absolute value is shown in the same (saturated) colorbar for all subplots.
    (a) The PFCMEq model with $\Delta t = 10\tau$ and (b) the sHPFC model with $\rho_0=\Gamma_S=2^{-6}$ at $t=4000\tau$.
    (c) and (d) show the same models, respectively, at $150\tau$ after the annihiliation event.
    A video showing the evolution of these fields (also considering other settings) is provided in the supplementary material \cite{supplementary_material}.}
    \label{fig:initial_state_relaxation}
\end{figure}

The main difference between these approaches lies in how the elastic field is treated.
Figure~\ref{fig:initial_state_relaxation} shows how the force density $\nabla \cdot h^{(\psi)} = \langle \tilde \mu_{\rm c} \nabla \psi - \nabla \tilde f \rangle$ evolves from initial state of Fig.~\ref{fig:benchmark_setup_and_PFCMEq} in the different models.
Far away from the dislocations, both the model allows for mechanical equilibrium, namely vanishing force density, with the PFCMEq retaining small oscillations owing to the two-step nature of the approach (evolution of the density field followed by a correction through $\vec u^{\delta}$) and its underlying assumptions.
Note that it enforces mechanical equilibrium similarly in bulk and at the core of dislocation, where deviations with respect to small deformations and bulk-like elastic constants are expected, thus introducing (minor) spurious effects.
The sHPFC, through its set of coupled equations and the damping term, allows for suppressing such oscillations. At the core, a residual driving force is observed, which reaches a stationary profile moving together with the dislocations.
Importantly, after the annihilation, the stress in the PFCMEq model is automatically set to zero by the algorithm, and no trace of the annihilation event is found, whereas the stress in the sHPFC model is relaxed through the excitation of elastic modes.
At these low values of $\rho_0$ and $\Gamma_S$, elastic equilibrium is attained almost immediately, but different scenarios may be obtained by varying parameters, as discussed in the following.

\subsection{Parameter space map}

\begin{figure}[htp]
    \centering
    \includegraphics[width=\textwidth]{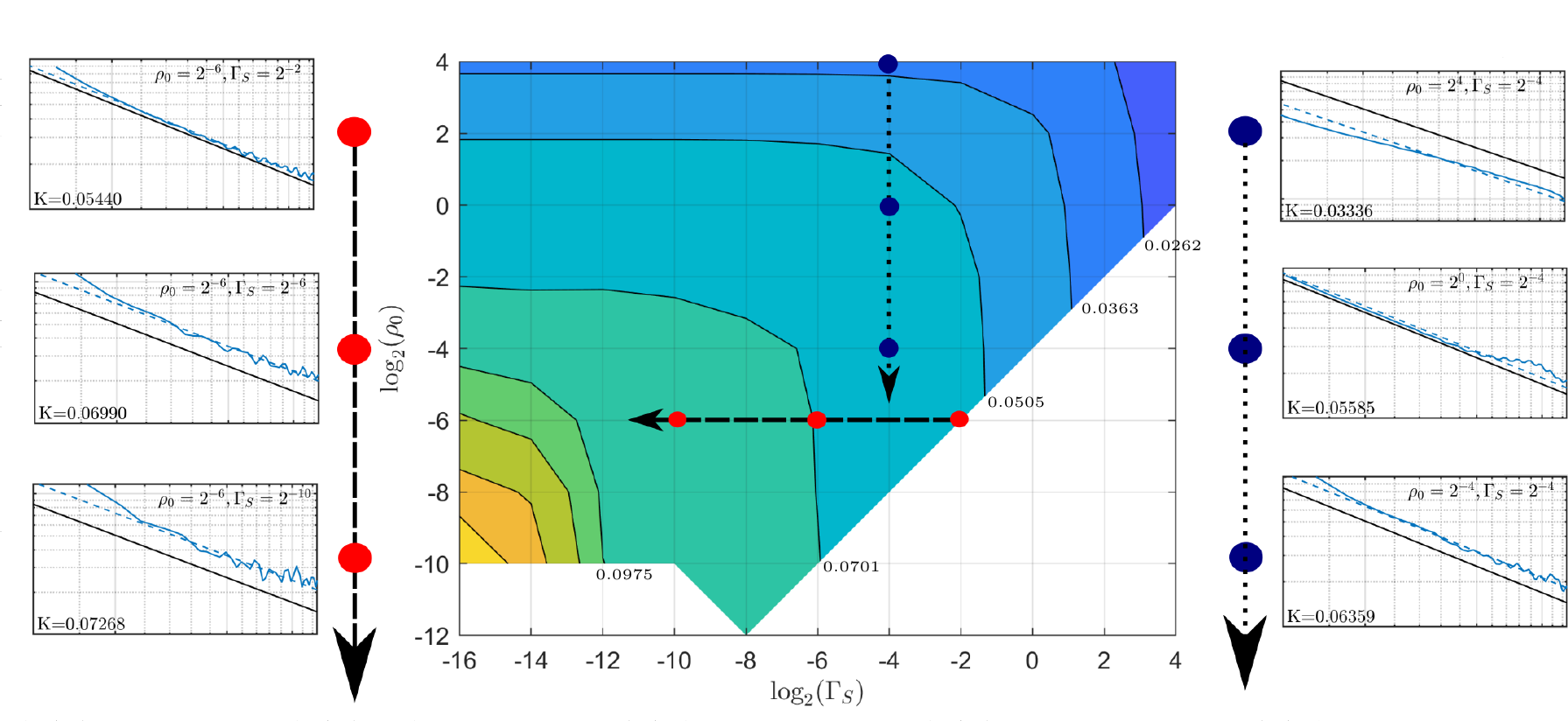}
    \caption{
    Contour plot of the dependence of the defect mobility $K$ on different parameters $\rho_0$ and $\Gamma_S$. The drawn lines correspond to representative isolines of $K$.
    For each grid point, the mobility $K$ was extracted by fitting the dependence of the defect speed on the dipole distance with $1/r$ as illustrated in the insets figures for few parameter values represented by red/blue dots.
    $K$, and thus the annihilation rate, increases with decreasing $\rho_0$ (blue points) and decreasing $\Gamma_S$ (red points).
    }
    \label{fig:parameter_search_figure}
\end{figure}

To explore the different dynamical regimes of the sHPFC model depending on the main parameters $\rho_0$ and $\Gamma_S$, we use the same setup of a dislocation dipole as in the previous section and vary systematically $\rho_0 \in (2^{-12},2^{-10},...,2^4)$ and $\Gamma_S \in (2^{-16},2^{-14},...,2^4)$, while keeping the rest of parameters fixed.
For each simulation  scanning the parameter space $(\Gamma_s,\rho_0)$, we determine the dislocation speed as function of the dipole size as shown in the inset of Fig.~\ref{fig:parameter_search_figure} for few parameter values.
We benchmark these graphs against the theoretical $1/r$ dependence from elasticity theory and determine mobility factor $K$ from a linear regression fit. The dependence of $K$ on the the varied parameters is plotted in Fig.~\ref{fig:parameter_search_figure}. It  shows the general trend that defect annihilation happens faster (larger mobility $K$) for smaller values of $\rho_0$ and $\Gamma_S$, which approaches the quasistatic limit.
For some parameter combinations, this induces numerical instabilities which prevented the evaluation of $K$ (uncolored points). As we focus here on the general trends, we adopted the same time discretization for every simulation, which is found to provide an accurate estimation of $K$ for most of its range. It is worth, however, noticing that for very small values of $\rho_0$ and $\Gamma_S$, enforcing a fast dynamic and thus delivering relatively large $K$ ($\gtrapprox 0.1$), an increased the temporal resolution should be considered to compute this quantity with similar accuracy.
Up to a global factor, a decreasing $\rho_0$ is equivalent to a decreasing $\Gamma$ responsible for the microscopic dissipative current $\tilde {\vec J}$.
The results obtained here are thus consistent with the small deformation limit found in the previous section. Namely, the model behaves like an elastic solid in this limit.
Additionally, the annihilation process also happens faster for a decreasing value of the dissipation parameter $\Gamma_S$.
This is consistent with other hydrodynamics models (e.g. Ref. \cite{heinonenConsistentHydrodynamicsPhase2016}) which have shown that the limit of elastic equilibrium is only attained in the limit $\Gamma_S\rightarrow 0$.

\section{Applications}

In this section, we report proof of concepts for the  applications of the sHPFC model, namely focusing on the dynamic of defects in three dimensions and polycrystalline systems.
To the authors' knowledge, these applications are beyond the capabilities of previously proposed PFC models including detailed modeling of elastic relaxation.

\subsection{Annihilation of loops}

In Ref. \cite{skogvollPhaseFieldCrystal2022}, a shear loop was initialized and it was discovered that its shrinkage followed a similar path under PFCMEq dynamics and PFC dynamics, but on a different time scale.
In this section, we study this process in more detail and also show that depending on the initial condition the loop shrinkage may take a very different path under sHPFC dynamics due to the interaction with the long-range elastic field.
The initial setup of the dislocation loop was prepared similarly to Ref. \cite{skogvollPhaseFieldCrystal2022} using parameters $(q_0,\Delta \texttt B_0,\texttt B^x_0,\texttt t,\texttt v,\psi_0)=(1,-0.3,1,0,1,-0.325)$ with $\mathcal L=\mathcal L_{1M}$, and $\rho_0=\Gamma_S=2^{-6}$ for the sHPFC model.

\begin{figure}
    \centering
    \includegraphics[]{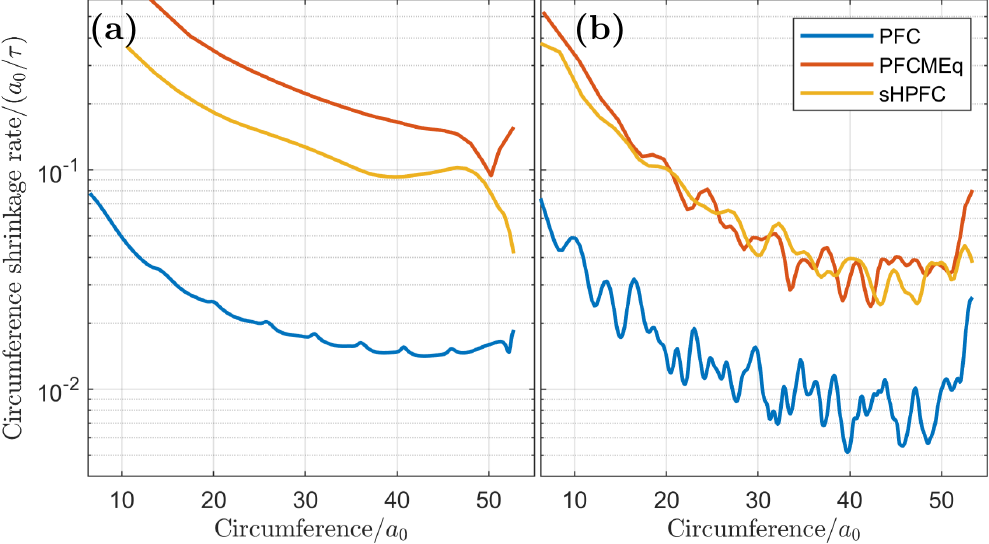}
    \caption{The circumference shrinkage rate for a dislocation loop with Burgers vector $\vec b = \frac{a_0}{2}[-1,1,1]$ seeded in (a) the (glide) plane $\vec n = (1,0,1)$ and (b) the plane $\vec n=(0,0,1)$. }
    \label{fig:shear_loop_annihilation_benchmark}
\end{figure}

\begin{figure}[htp]
    \centering
    \includegraphics[width=\textwidth]{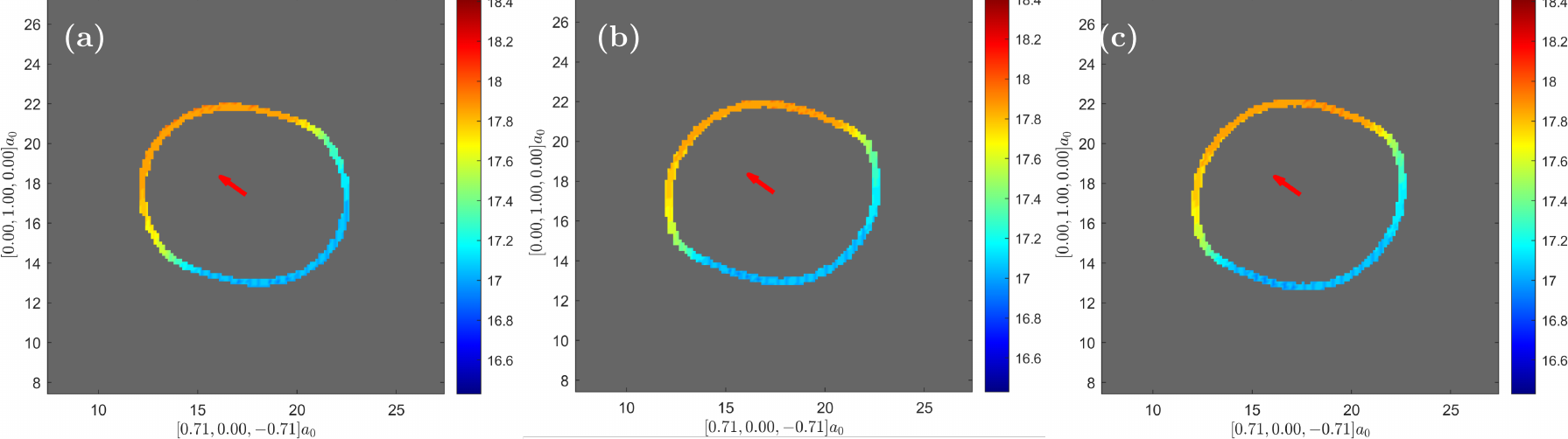}
    \caption{The dislocation loop with Burgers vector $\vec b = \frac{a_0}{2}[-1,1,1]$ (red quiver), initially seeded in the $\vec n = (1,0,1)$ -(glide)plane after shrinking to a circumference of $l_\mathcal C \approx 31 a_0$.
    It is shown for (a) the PFC model at $t=1420\tau$, (b) the PFCMEq model with $\Delta t = 0.1 \tau$ at $t=140\tau$ and (c) the sHPFC model with $\rho_0=\Gamma_S=2^{-6}$ at $t=230\tau$.
    The first and second axis indicate, respectively, the two orthonormal directions in the initial loop plane.
    The colorbar indicates the loop's vertical position relative to this plane.
    A video of this annihilation event can be found in the supplementary material \cite{supplementary_material}.}
    \label{fig:Loops101_images}
\end{figure}

The initial loop was first inserted in the $\vec n=(1,0,1)$ plane and had a slip (Burgers vector) of $\vec b = \frac{a_0}{2}[-1,1,1]$, which is the most common slip direction in bcc materials \cite{weinbergerSlipPlanesBcc2013}.
The resulting shrinkage rate $|\partial_t L|$ of the dislocation loop having length $L$ is shown in Fig.~\ref{fig:shear_loop_annihilation_benchmark}.
The loop shrinks similarly in all three models thus far considered (PFC, PFCMEq, and sHPFC), but on different time scales.
Fig.~\ref{fig:Loops101_images} shows the similar loops in the reference plane as obtained with the three methods at a given time during shrinkage. They evolve in all cases without leaving the initial glide plane in which they were initialized.

\begin{figure}[htp]
    \centering
    \includegraphics[width=\textwidth]{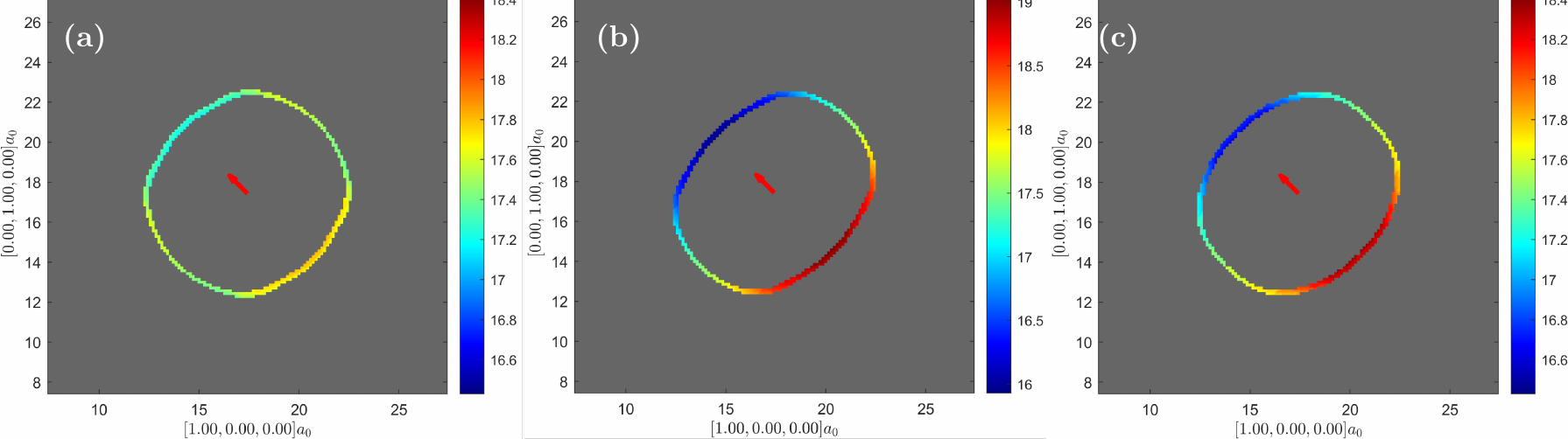}
    \caption{The dislocation loop with Burgers vector $\vec b = \frac{a_0}{2}[-1,1,1]$ (red quiver, projected), initially seeded in the $\vec n = (0,0,1)$-plane after shrinking to a circumference of $l_\mathcal C \approx 31 a_0$.
    It is shown for (a) the PFC model at $t=2610\tau$, (b) the PFCMEq model with $\Delta t = 0.1 \tau$ at $t=600\tau$ and (c) the sHPFC model with $\rho_0=\Gamma_S=2^{-6}$ at $t=630\tau$.
    The first and second axis indicate respectively the two orthonormal directions in the initial loop plane.
    The colorbar indicates the loop's vertical position relative to this plane.
    A video of this annihilation event can be found in the supplementary material \cite{supplementary_material}. }
    \label{fig:Loops001_images}
\end{figure}

To check whether the same holds for a loop of less simple topology, we also considered a mixed type (neither shear nor prismatic) dislocation loop with the same Burgers vector but in a different initial plane, $\vec n=(0,0,1)$.
The results are shown in Fig.~\ref{fig:Loops001_images}.
In this case, the trajectory is much more dependent on the particular dynamics.
Notably, from Fig.~\ref{fig:Loops001_images}, we see that for a loop with such a mixed nature, under purely classical PFC dynamics, the loop does not move out of plane.
For the models that incorporate elastic effects, however, we see that the loops move out of the initial plane and conclude that this is due to the long-range elastic interactions.

\subsection{Multi-grain network at mechanical equilibrium}

In the examples considered so far, we have looked at a single dislocation dipole or loop in an otherwise perfect lattice.
In such cases, the problem of solving for elastic equilibrium can be done by treating the whole system as an elastic medium and assuming that the elastic constants are unaffected by the presence of defects.
In the case of a highly deformed solid, or a polycrystal with multiple crystallographic orientations, the elastic problem can not be readily solved by the methods proposed in the PFCMEq model.
This is particularly relevant in the case of crystal structures featuring anisotropic elasticity, which includes the square phase shown in Fig.~\ref{fig:MicroStructures} and most 3D PFC crystal phases \cite{skogvollStressOrderedSystems2021}.
The hydrodynamic treatment of the sHPFC model, however, knows no such limitations. To highlight this, we compare the residual stress produced in this model with the state of stress in the classical PFC model.

\begin{figure}[htp]
    \centering
    \includegraphics[width=\textwidth]{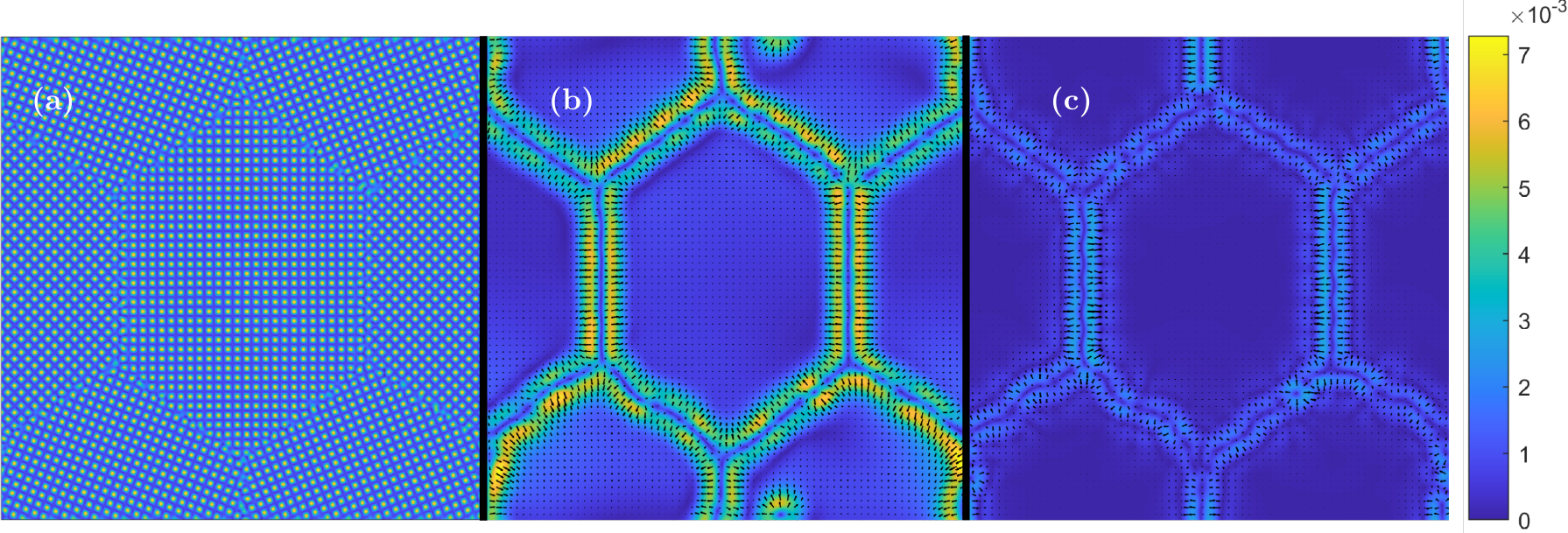}
    \caption{(a) A polycrystalline square PFC configuration is evolved for $1000\tau$ and the force density $\nabla \cdot h^{(\psi)} = \langle \tilde \mu_{\rm c} \nabla \psi - \nabla \tilde f \rangle$ is shown at the end of the simulation when evolved according to (b) PFC dynamics and (c) sHPFC dynamics.
    The colorbar, in units of $\mu/a_0$, is the same for panels (b) and (c). }
    \label{fig:square_polycrystal}
\end{figure}

The question of whether long-range elastic effects play an essential role in the stress state of polycrystalline configurations, or whether grain boundaries provide a sufficient screening both for interfacial and bulk stresses is a subject of debate.
In order to investigate this issue, we prepare a 2D polycrystalline aggregate composed of square lattices (Fig.~\ref{fig:square_polycrystal}), using parameters  $(q_0,\Delta \texttt B_0,\texttt B^x_0,\texttt t,\texttt v,\psi_0)=(1,-0.3,1,0,1,-0.3)$ and $\mathcal L = \mathcal L_{2M}$, and evolve it according to the PFC and sHPFC dynamics, respectively, for $1000\tau$.
The force density $\nabla \cdot h^{(\psi)}$ at the end of the simulation is shown in Figs. \ref{fig:square_polycrystal}(b) and (c), respectively.
We see that in the classical PFC there is still residual stress both in the bulk and at grain boundaries, where residual stress is particularly large and thus influences their kinetic evolution.
This is consistent with the results for the hydrodynamic model in the APFC framework \cite{heinonenPhasefieldcrystalModelsMechanical2014}, which showed that the shrinkage of a rotated inclusion happened much faster when accounting for elastic equilibrium.

The simulation setup used for the 2D square PFC in this section is the 2D analog of the bcc PFC setup study of diffusion-mediated plasticity in the stress-strain tests of Ref. \cite{berryAtomisticStudyDiffusionmediated2015}, in which integer creep law exponents were found in the Nabarro-Herring creep regime.
Figure \ref{fig:square_polycrystal} indicates that the sHPFC model may be suited to model plastic deformation in the higher strain regime where deformation is  mediated by dislocations.

\section{Conclusion}

In summary, we have provided a systematic derivation of how the evolution of a density field with a microscopic structure can be coupled with a hydrodynamic velocity field, using the principle of free energy minimization.
The key insight of the derivation comes from systematic use of spatial coarse-graining, which has also previously been exploited in other hydrodynamic approaches.
The resulting model has several key features that have been missing from previous modeling approaches, such as the incorporation of a spatially dependent mobility, which inhibits mass flux by diffusion in the homogeneous liquid phase, as well as a precise definition of the stress field acting on the density that includes both reversible and irreversible components.
Using the phase field crystal free energy as a prototypical model to investigate the formalism, we have tested the framework in two limits.

First, at zero wave dynamics, we observe that during the propagation of a liquid-structure interface during solidification in a super-cooled liquid, the type of mobility used in diffusion plays a central role.
Isotropic mobility causes a more diffuse interface between the phases and all phases to grow equally fast in all directions, which for an anisotropic microscopic structure (such as stripes) produces interference patterns on the scale of the microscopic structure itself.
With the spatial mobility modification, however, we see that the anisotropy of the microscopic structure is maintained, which produced an anisotropic growth rate for the simulation parameters used in this study.

Second, using the approximation of a constant average density, we derive a simplified hydrodynamic phase field model (sHPFC), which in the low density and dissipation limit is shown to reproduce correct dislocation dynamics.
This formalism is used to extend the investigation of shrinkage of a dislocation loop done in Ref. \cite{skogvollPhaseFieldCrystal2022}.
It is shown that while classical PFC dynamics give only qualitatively the correct behavior for a simple shear loop annihilation, elastic interactions create a different kinematic path for loop annihilation for more complicated initial conditions.
This has wide-reaching consequences for PFC modeling of dislocation dynamics in three dimensions since this shows that elastic interactions of dislocations must be taken into account.
Finally, we exploit the fact that the sHPFC does not need a reference lattice configuration to evolve a poly-crystalline sample where the long-range elastic fields are properly related.

In this paper, we have focused on applications close to the quasistatic limit, i.e. keeping the PFC at mechanical equilibrium.
However, as the model explicitly handles the dynamics of elastic waves, it is also suitable for studying systems where elastic effects happen on a timescale comparable to that of defect motion, such as annihilation events, nucleation, or fracture.

\section*{Acknowledgments}
We thank Jorge Vi\~nals for many stimulating discussions and his valuable comments at various stages during this work. MS acknowledges support from the German Research Foundation (DFG) under Grant No.~SA4032/2-1.

\section*{Reference}

\bibliographystyle{unsrt}
\bibliography{Bibliography}

\end{document}